\documentclass[12pt,preprint]{aastex}
\shorttitle{Precessing Jet Excavating a Protostellar Envelope}
\shortauthors{Ybarra et al.} 
\begin{document}
\title{First Evidence of a Precessing Jet Excavating a Protostellar Envelope}
%
\author{Jason E. Ybarra\altaffilmark{1} \& Mary Barsony\altaffilmark{1,2}}
\affil{Department of Physics \& Astronomy, San Francisco State University\\
1600 Holloway Drive, San Francisco, CA  94132}
\email{jybarra@stars.sfsu.edu, mbarsony@stars.sfsu.edu}
%
%
\author{Karl E. Haisch, Jr.\altaffilmark{1}}
\affil{Utah Valley State College, 800 West University Parkway, Orem, UT 84058}
\email{haischka@uvsc.edu}

\author{Thomas H. Jarrett\altaffilmark{1}}
\affil{Infrared Processing and Analysis Center, MS 100-22 Caltech, Pasadena, CA 91125}
\email{jarrett@ipac.caltech.edu}

\author{Raghvendra Sahai\altaffilmark{1}}
\affil{Jet Propulsion Laboratory, MS 183-900, 4800 Oak Grove Drive, Pasadena, CA 91109}
\email{Raghvendra.Sahai@jpl.nasa.gov}

 \and
   
 \author{Alycia J. Weinberger}
 \affil{Department of Terrestrial Magnetism, Carnegie Institution of Washington, Washington, DC 20015}
 \email{weinberger@dtm.ciw.edu}

\altaffiltext{1}{Observations with the Palomar 5 m telescope were obtained under 
a collaborative agreement between Palomar Observatory and the Jet Propulsion Laboratory.}


\altaffiltext{2}{and Space Science Institute, 
4750 Walnut Street, Suite 205, Boulder, CO  80301}

\begin{abstract}
We present new, sensitive, near-infrared images of the Class I protostar, Elias 29,
in the Ophiuchus cloud core.   To explore the relationship between the infall envelope and the outflow, narrowband H$_2$ 1$-$0 S(1), Br$\gamma$, and $K_{\rm{cont}}$ filters 
were used to image the source with the Wide-Field Infrared Camera on the Hale 5.0-meter
telescope and with Persson's Auxiliary Nasmyth Infrared Camera on the Baade 6.5-meter telescope.
The source appears as a bipolar, scattered light nebula, with
a wide opening angle in all filters, as is typical for late-stage protostars. However, the pure H$_2$ emission-line images point to the presence of a heretofore undetected precessing jet.  It is argued that high velocity, narrow, precessing jets provide the mechanism for creating the observed wide-angled outflow cavity in this source.
\end{abstract}


\section{Introduction}

Powerful, bipolar supersonic outflows are a hallmark of the earliest 
protostellar stages.  
Molecular outflows generally exhibit wide opening angles and
velocities of $\sim$ 10--30 km s$^{-1}$.
Optical jets powered by protostars
\textit{}
exhibit small opening angles and velocities of $\sim$ 100-400 km s$^{-1}$
\cite{mun90,fri05}.   Due to the disparate spatial and velocity distributions of the molecular and optical outflow components, a two-wind structure has often been invoked to explain the instances in which
both outflow components are observed to be powered by a single source \cite{sto88,dav02,lis05}.
However, an alternative scenario, in which optical jets
and molecular outflows result from a single, collimated flow is
gaining wide acceptance \cite{har94, san99, wol00, yun01}.

An outstanding, unsolved problem with jet-driven outflow models
is how jets produce the wide-angle outflow
cavities observed in both protostellar envelopes and in molecular outflows.
A proposed solution is that of jets changing direction \cite{mas93}, for which there is now 
growing evidence \cite{brp96, rei97, bar98, arc05}.
Nevertheless, the spatial relationship between the outflow cavity and the currently active wandering jet has not yet been directly observed within the boundaries of a protostellar envelope.
We use sensitive, narrowband near-infrared imaging to study this relationship: 
the scattered light emission from the protostellar envelope is imaged with
a narrowband continuum filter, whereas the currently active jet regions within the infall envelope
are traced in continuum-subtracted, narrowband H$_2$ images, sensitive to shock-excited emission.
Based on such observations of the Class I protostar, Elias 29$=$WL15 ($\alpha_{2000}=16^h27^m09.43^s$, $\delta_{2000}=-24^{\circ}37^{\prime}18\farcs7$) in the nearby (d$=$125 pc) $\rho$ Ophiuchi molecular cloud, we report the first detection of a precessing jet carving out a protostellar envelope's cavity.

\section{Observations and Data Reduction}

Near-infrared, narrowband imaging of Elias 29 was undertaken at two facilities, Palomar and Las Campanas
Observatories.  The Wide Field Infrared Camera (WIRC) on the Hale 5.0-meter
telescope at Palomar Observatory, with its 8\farcm 7
$\times$ 8\farcm 7 field of view, was used to image the field including Elias 29  \cite{wil03}.  
WIRC employs a 2048$\times$2048 pixel Hawaii-II detector with a
plate-scale of  0\farcs 2487 pixel$^{-1}$.  The K-band seeing was $\sim$ 1\farcs 0
on the night of the observations, 2004 July 11.
Two narrowband filters were used for imaging:
an H$_{2}$ 1--0 S(1) filter centered at 2.120 $\mu$m with a 1.5\% bandpass
and a line-free continuum filter (K$_{\rm{cont}}$) centered at 2.270 $\mu$m with a 1.7\% bandpass.
%
The telescope was dithered in a nine-point pattern, with 6$^{\prime\prime}$
offsets between dither positions.
The integration time at each dither position was 60 sec (30 sec x 2 coadds).
The dither pattern was repeated, after a 5$^{\prime\prime}$ offset of the telescope, until
a total integration time of 27 minutes was reached in each filter.
%
%
Elias 29 was re-observed with Persson's Auxiliary Nasmyth Infrared Camera (PANIC)
on the Baade 6.5-meter telescope at Las Campanas Observatory (LCO) on the night of
2005 June 17 UT. 
The K-band seeing was 0\farcs 5.  PANIC employs a 1024$\times$1024 Hawaii-I detector 
with a plate scale of  0\farcs 125 pixel$^{-1}$ \cite{mar04}.   The two narrowband filters used for imaging at LCO
were the H$_{2}$ 1--0 S(1) filter centered at 2.125 $\mu$m with a 1.1\% bandpass,
and the Br$\gamma$ filter, centered at a wavelength of 2.165 $\mu$m with a 1.0\% 
bandpass.  The Br$\gamma$ filter served as a surrogate narrowband continuum filter for
the PANIC dataset, in the absence of a line-free, narrowband continuum filter adjacent
to the  H$_{2}$ filter at LCO.
%
The observing pattern for Elias 29 at LCO consisted of interleaved on- and off-source integrations, of 30 sec duration each, through each filter.  
On-source integrations were dithered in a checkerboard pattern,
with 12$^{\prime\prime}$ offsets. Five-point dithered sky observations were taken in a clear patch  3$^{\prime}$ E of Elias 29.
%
%
The final on-source integration times with PANIC were 70 mins. through the
H$_2$ filter and 67.5 mins. through the Br$\gamma$ filter.

All data were reduced using IRAF\footnotemark[3]. 
Object frames were sky-subtracted, 
flat-fielded, and corrected for bad pixels for each night's observations through each narrowband filter.  
Individual processed images for each filter/instrument combination were then aligned using the available point sources in common with the IRAF task IMCENTROID, before combining to produce final images. 

Although the H$_2$ filters are so-called because they transmit the 2.12$\mu$m emission line, they both also transmit some continuum emission, if present in the source. Therefore, to to trace the morphology of any outflow/jet component within the protostellar envelope of Elias 29, the bright continuum emission from the scattering envelope must be subtracted from the H$_2$ filter images. This requires proper scaling and alignment of the narrowband ``continuum'' images before subtracting them from the respective H$_2$ filter images. These  tasks were achieved
separately for the WIRC and PANIC data.  
For the PANIC images, shown in Figure 1, the Br$\gamma$ filter image was scaled by a factor of 0.82, which is simply the ratio of the transmission of the H$_2$ filter to that of the Br$\gamma$ filter,  from the filter transmission curves.
Proper alignment of the H$_2$ and scaled ``continuum'' images before subtraction,
crucial for obtaining accurate pure H$_2$ line-emission images, was achieved
by using the IRAF task, IMCENTROID, for common point sources in each filter.

For astrometry, a plate solution was determined for the WIRC data by using a least squares fit for 7 point-like objects in common with 2MASS \footnotemark[4]. The 2MASS reference objects had uncertainty ellipses of $\sim$ 0\farcs09. The plate solution had a 0\farcs2 RMS residual, our estimated astrometric uncertainty.

\footnotetext[3]{\textbf{IRAF} is distributed by the National Optical Astronomy Observatories, which are operated by the Association of Universities for Research in Astronomy, Inc., under cooperative agreement with the National Science Foundation.}
\footnotetext[4]{http://www.phys.vt.edu/$\sim$jhs/SIPbeta/astrometrycalc.html}

\section{Results and Discussion}

The reduced WIRC and PANIC images of Elias 29 appear essentially identical. 
Only the PANIC images are presented here, since these were obtained in better seeing and have higher signal-to-noise.  Figure 1 shows our results,
all presented at a linear greyscale stretch to emphasize the presence of the
pure emission features, 3a and 3b in Figure 1a, and their absence from Figure 1b. Figure 1a shows the appearance of Elias 29
through the narrowband H$_2$  filter.
Figure 1b shows Elias 29 through the surrogate ``narrowband continuum'' (Br $\gamma$) filter.
Since the NIR continuum emission overwhelms the strength of the pure H$_2$ line emission for this object, Elias 29
looks similar in both filters.
Figure 1c shows the continuum-subtracted, pure H$_2$ emission-line image of Elias 29.
Saturation effects from the extreme brightness of Elias 29 prevent correct image subtraction within the central $\sim$ 5\farcs5 radius region, resulting in
the black and white circular artifacts at the center of Figure 1c. Outside this area,
five regions of pure H$_2$ line emission are discovered, having been 
missed by previous investigators \cite{gom03, kha04}.
%
%
The appearance of Elias 29 in Figures 1a \& 1b is
consistent with scattered light models of Class I protostars consisting of an
infall envelope containing a bipolar cavity \cite{whi03}.
The boundary of the wide-angled cavity walls within the protostellar infall envelope
is indicated by the lowest contour levels overplotted on Figures 1b \& 1c.
The lack of a narrow ``waist'' or distinctive hourglass shape at this angular
resolution is consistent with a disk of small spatial extent (B. Whitney, priv. comm.).
%
%
%
Elias 29 appears strikingly different in the continuum-subtracted, pure H$_2$ emission line image than in the narrowband filter images.
The pure H$_2$ emission objects, labelled 1, 2a, 2b, 3a, and 3b, appear
at the same location, and have similar morphologies in both the PANIC and the WIRC
(not shown here) continuum-subtracted images.  Taken 
together, the H$_2$ emission knots form a sinuous structure with S-shaped point-symmetry about the center.

Table 1 lists the observed characteristics of each H$_2$ feature.
The first column lists the feature designations from Figure 1c. 
The second and third columns list the coordinates of each H$_2$ emission region,
corresponding to the WIRC pixel with the highest counts in each object.
The coordinates are derived from the WIRC data,
since there were not enough sources in the smaller PANIC
field to provide a plate solution for absolute astrometry.
The final, continuum-subtracted PANIC image was used to determine
the signal to noise ratio of each emission feature, listed in the last column of Table 1,
since the seeing and signal to noise were better in the PANIC than in the WIRC dataset.
 A rectangular aperture (of dimensions given in the fourth column of Table 1) was used to determine the average counts in each feature. 
Several background regions were used to calculate the average background noise level and its standard deviation.  The backgound level was subtracted from the average counts in the rectangular aperture containing each H$_2$ emission region, and the result was divided by the standard deviation of the background to give the SNR. 
Only features 3a and 3b have $\sigma$ $<$ 3. Nevertheless, these objects are real, since they appear at identical locations and have similar morphologies in both datasets (WIRC  and PANIC).
%
%

H$_2$ emission associated with molecular outflows generally arises
from shock interactions of the unseen, fast jet/wind component with the
ambient medium.  Elias 29 is known to drive a CO outflow \cite{bon96, sek97}, with velocities as high as $\sim$ 80 km s$^{-1}$ \cite{boo02}, although
its morphology is unknown at the scales of the images presented here.
In some Class I protostellar envelopes, both
spatially diffuse H$_2$ emission and compact H$_2$ emission knots have been 
identified in the same source \cite{yam92, dav02}.  In such cases, 
the diffuse component is found to
trace the infrared continuum reflection nebula, whereas 
knotty, compact emission traces the fast jets observed at optical
or cm-continuum wavelengths. By analogy, the knotty H$_2$ emission
features identified in Figure 1c are likely to be associated with 
the jets driven by Elias 29.  Since the cooling time of shocked 2.12 $\mu$m
H$_2$ emission is of order $\sim$ a few years \cite{shu82,all93}, the newly identified 
features in Figure 1c are tracing recent shock activity.
%
%
%
The three H$_2$ emission features in Figure 1c: 1, 2a, and 2b,
describe an S-shaped, point-symmetry, about the center, consistent with a precessing jet. Features 3a and 3b are extended and diffuse, close to the inferred cavity wall within the southern hemisphere of the protostellar envelope of Elias 29.  
More quantitative results on the precessing jet, such as
its precession period and initial opening angle 
await HST NICMOS polarimetry. Ground-based, AO-assisted  integral field spectroscopy
of the H$_2$ line would establish the velocity field.
%

Hydrodynamic simulations of molecular outflows driven by pulsed, precessing protostellar
jets have become available only recently \cite{lim01,ros04, smi05}. One stated aim of these studies is to model the outflows to infer properties of the (unseen) driving jets, such as jet power and precession.  Currently, only a limited set of models have been 
calculated with accompanying simulated H$_2$ images.  Assumed jet radii
are 1.7 $\times$ 10$^{15}$ cm, corresponding to $\sim1\farcs0$
at the distance to Elias 29 (so, well beyond the disk region where
jets may be generated). The pulsation is input as a 30\% velocity modulation with a 60 yr period into the models.
The timespan over which the models are calculated
is short (t $<$ 500 yr), but much longer than the dynamical time of a few decades
for a jet travelling a conservative $\sim$ 100 km s$^{-1}$ to reach the H$_2$ 
Features 1 and 2a in Elias 29. This, compared with the very fast cooling time of H$_2$,
validates comparison of the Elias 29 pure emission-line H$_2$ image (Figure 1c)
with simulated H$_2$ images. 

Models of molecular jets with precession angles of 5$^{\circ}$, 10$^{\circ}$, and 20$^{\circ}$ have been calculated, although the precession angle of the Elias 29 jet
may well be even wider, based on the projected half-angles formed by Features 1 and 2a with the central symmetry axis of the observed reflection nebulosity.  Simulated H$_2$ images from two general classes of precessing jet models have been published: Jets with assumed precession periods of 50 yr, termed ``fast'' precessing,
and jets with precession periods of 400 yr, termed ``slow'' precessing \cite{ros04,smi05}.  The terms ``fast'' and ``slow'' here are relative to the flow evolution time of 500 yr.
Comparison of Figure 1c with synthetic H$_2$ $v=1-0$ S(1) model images from such simulations leads to the conclusion that this jet exhibits slow, rather than fast, precession,
that is to say, qualitatively, Figure 1c resembles the H$_2$ 1$\rightarrow$0 morphologies
presented in Figures 6 \& 7, rather than those of Figure 5, of Smith \& Rosen (2005).
For slow precession, the simulations show ordered chains of bow shocks and meandering streamers of H$_2$, which contrast with the chaotic H$_2$ structures produced by jets in rapid precession. The production of specific simulated
H$_2$ images for direct comparison with Figure 1c is beyond the scope of this paper,
but may be attempted in the future.
%

Images of both scattered light cavities and H$_2$ jet emission within them
exist for only a handful of Class I objects \cite{dav02}.  For all of these,
the jets lie well within the outflow cavity walls, close to, if not precisely along,
the symmetry axis of each outflow cavity. By contrast, the jet of Elias 29, as traced
by H$_2$, is found close to, and, in one case (Feature 2b) along, the cavity walls (Fig. 1c).

Examples of other S-shaped, precessing outflows are known (Gueth et al. 1996;
Schwartz \& Greene 1999; Hodapp et al. 2005; Sahai et al. 2005).
However, no precessing protostellar jets have been imaged at their base within their scattered light cavities, either because they are too
embedded to be detected in the NIR (L1157, HH211-mm), or because of lack of narrowband continuum data excluding the H$_2$ emission line for proper image subtraction(IRAS 03256$+$3055).Binarity has inevitably been invoked to account for the presence of
precessing jets \cite{ter99}.  However, Elias 29 is known to be single \cite{sim95, ghe93}. At scales between $\sim 1-60$AU, lunar occultation observations show
that 90\% of the K-band flux from Elias 29 originates in a component 
$\sim$ 7 mas diameter ($\sim$ 0.8 AU) with the remaining 10\% of the K-band
flux originating in a much larger, 415 mas diameter, diffuse component \cite{sim87}.
It is then inferred that within $\approx$ 0.8 AU, Elias 29 is a single object. 
Alternative scenarios leading to jet precession that may be operative in Elias 29
might be disk oscillations associated with 
a propeller-driven outflow, recently discovered in MHD simulations
\cite{rom06} or warping of the accretion disk caused by the 
magnetically driven jets \cite{lai03}. Future high-resolution integral field spectroscopic observations of both the jet and the disk components could test the latter model, which predicts the disk and jets to precess in opposite senses.

We thank our very professional night assistants, Jean Mueller at Palomar Observatory
and Hernan Nu\~nez at Las Campanas Observatory.
Portions of this research were funded by NSF AST-0206146 to M.B. and by the 
NASA LTSA Program (NAG5-8933).  NASA/LTSA Award 399-30-61-00-00 to R.S. is
gratefully acknowledged.   AJW acknowledges support from NASA Space
Science grant NAG5-3175.

\clearpage

\begin{deluxetable}{lllrr}
\tablewidth{0pt}
\tablecaption{H$_2$ emission features}
\tablehead{
\colhead{H$_2$ Feature} & \colhead{$\alpha_{2000}$ } & \colhead{$\delta_{2000}$} & 
\colhead{Aperture} & \colhead{S/N}\\
\colhead{Designation} & \colhead{ (h$\ \ $m$\ \ $s$\ \ $)} & \colhead{ \arcdeg$\ \ $  \arcmin$\ \ ^{\prime\prime}\ \ $ } & 
\colhead{} & \colhead{} 
}
\startdata
 1 & 16 27 08.79 & -24  37 16.5 & $0\farcs6 \times 0\farcs5$ & 3.5  \\
 2a & 16 27 10.04 & -24 37 20.8 & $0\farcs6 \times 0\farcs5$ & 3.9 \\
 2b & 16 27 10.70 & -24 37 24.3 & $1\farcs3 \times 0\farcs5$ & 14.2 \\
 3a & 16 27 11.64 & -24 37 56.1 &  $0\farcs6 \times 0\farcs5$ &2.6 \\
 3b & 16 27 11.88 & -24 38 00.0 & $0\farcs9 \times 0\farcs4$ & 2.8 \\
\enddata

\end{deluxetable}

\clearpage

\begin{figure}
\plotone{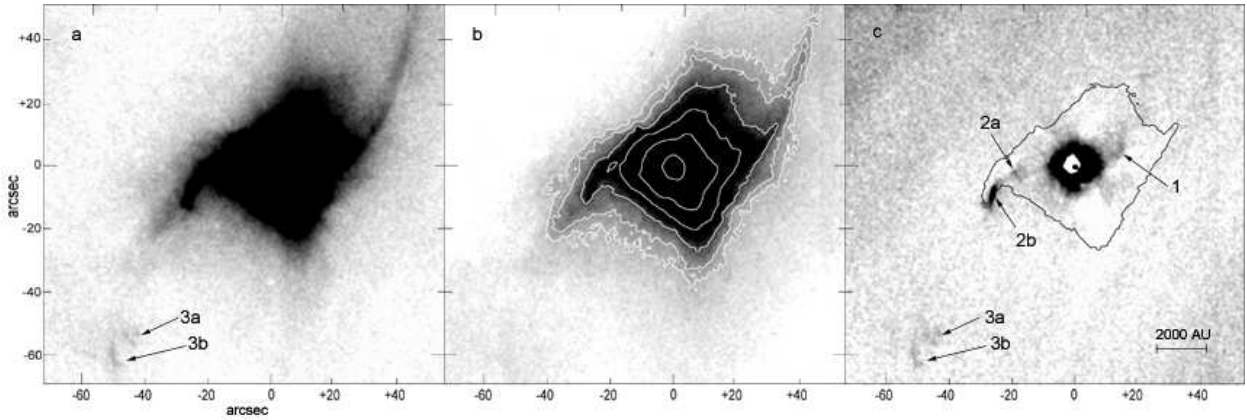}
\caption{Narrowband Images of Elias 29. N is up and E is left in all panels. Offsets from RA(2000)= $16^h27^m09.43^s$ and Dec(2000)= $-24^{\circ}37^{\prime}18\farcs7$ are indicated, in arcsec units.
(a) Elias 29, observed through the narrowband H$_2$ filter with PANIC, (b) Elias 29 imaged with the Br$\gamma$ filter with PANIC, contour levels are 3, 5, 10, 20, 50, 500 $\sigma$. Note the absence of Features 3a and 3b in the Br$\gamma$ image, and their presence in the H$_2$ filter image. (c) Pure H$_2$ emission-line image of Elias 29, obtained by subtracting the scaled and aligned Br$\gamma$ filter image from the PANIC H$_2$ filter image with the Br$\gamma$ 10 $\sigma$ contour superimposed. The pure H$_2$ emission line features discussed in the text and listed in Table 1 are labelled.}
\label{Figure 1}
\end{figure}

\end{document}